\documentclass[12pt,titlepage]{article}
\usepackage{epsfig}
\usepackage{graphicx}
\usepackage{amsmath}

\def\beq{\begin{equation}}
\def\eeq{\end{equation}}
\def\bea{\begin{eqnarray}}
\def\eea{\end{eqnarray}}
\def\bean{\begin{eqnarray*}}
\def\eean{\end{eqnarray*}}

\def\p{\tilde p}
\def\s{\tilde s}
\def\t{\tilde t}

\begin{document}

\title{Vacuum energy and the latent heat of AdS-Kerr black holes}

\author{{Brian P. Dolan}\\
{\em Department of Mathematics, Heriot-Watt University}\\ 
{\em Colin Maclaurin Building, Riccarton, Edinburgh, EH14 4AS, U.K.}\\
[5mm]
{\em Maxwell Institute for Mathematical Sciences, Edinburgh, U.K.}\\
[5mm]
{\em Email:} {\tt B.P.Dolan@hw.ac.uk}
}

\maketitle
\begin{abstract}
Phase transitions for rotating asymptotically anti-de Sitter black holes
in four dimensions are described in the $P-T$ plane, in terms of the Hawking temperature and the pressure provided by the cosmological constant. 
The difference between constant angular momentum and constant angular velocity
is highlighted, the former has a second order phase transition while the
latter does not.
If the angular momentum is fixed there a line of first order phase transitions terminating at a critical point with a second order phase transition and vanishing latent heat,
while if the angular velocity is fixed there is a line of first order phase transitions terminating at a critical point with infinite latent heat.
For constant angular velocity the analytic form of the 
phase boundary is determined, latent heats derived and the Clapeyron 
equation verified.

\bigskip

\bigskip
\noindent

\begin{flushright}
\end{flushright}

\end{abstract}

\section{Introduction}

There has recently been interest in interpreting the
cosmological constant in an asymptotically de Sitter or anti-de Sitter 
black hole space-time to be a pressure in a thermodynamic sense.
The thermodynamically conjugate variable would then be interpreted as a
volume associated to the black hole, though it would not necessarily be
related to any notion of geometric volume, \cite{KRT} \cite{BPD1} \cite{CGKP}.
The idea of varying the cosmological constant goes back to 
\cite{HenneauxTeitelboim} \cite{Teitelboim}, but the interpretation
adopted here was first given in \cite{KRT}.
In this paper we re-examine some of the known thermodynamics of asymptotically anti-de sitter (AdS) Kerr
black holes from this perspective, emphasising the role that the pressure 
plays in the analysis.

As usual in thermodynamics, the phase structure depends on the constraints.
For example when the angular momentum $J$ is held fixed there is a 
critical point with a second order phase transition at finite pressure and temperature, first found in \cite{CCK}. This
constant $J$ phase transition is known to have mean field exponents, 
\cite{GKM} \cite{PdV}, putting it in the same universality class as a van der Waals gas and the phase diagram (figure \ref{fig:J_free_energy}) looks the same as that of a van der Waals gas. 

On the other hand 
fixing the angular velocity $\Omega$ results in a phase diagram
which, like the one dimensional Ising model, has no second order transition at any finite temperature.  
There is a critical point at finite $T$ and $P$ where the free energy has a cusp, but 
the latent heat diverges there.
Strictly speaking there is a second critical point, with vanishing latent
heat, but it is at infinite $T$ and there is no phase transition as one cannot pass through this point ---  this is similar to the 
$1$-d Ising model, though there the critical point is at $T=0$.

The phase structure for constant $\Omega$
and constant $J$ are different in the
the pressure-temperature plane.\footnote{The fixed $J$ and fixed $\Omega$ ensembles were  
analysed in \cite{BMS} and \cite{BKR} in terms Ehrenfest equations.}  In the former case the 
phase boundary is determined by the condition $\rho=2P$, where $\rho=\frac{M}{V}$ is the black hole mass per unit volume, while it is not so easy find
an analytic expression for constant  $J$.
Many familiar notions from ordinary thermodynamics are
applicable, such as the Clapeyron equation for the slope of the phase boundary
in the $P-T$ plane, but there are also significant differences.

While the analysis produces some explicit expressions for phase boundary curves
in the $P-T$ plane and latent heats in black hole phase transitions that have not appeared in the literature before, these
are not the main point of the paper, being just trivial consequences
of the structure of the Hawking-Page phase transitions.  Rather the main
point is to emphasise the shift in viewpoint that occurs when the thermodynamic
volume is introduced. The phase diagram in figure \ref{fig:Omega_free_energy} the same
as that of \cite{HHT-R}, but drawn using thermodynamic variables rather than the geometric variables that are explicit in the metric.
This is done to emphasis the physics of the thermodynamics: the free energy
is a single valued function of the geometric variables, but is multiple valued
in terms of thermodynamic variables, giving the different branches in the $P-T$ plane that are the hallmark of phase transitions in the grand canonical ensemble.
This conceptual shift may well prove to 
be important for the analysis of rotating superfluids in the AdS/CFT correspondence \cite{Brihaye} 

In section \S\ref{sec:Static}, static non-rotating black holes are
treated, as a warm up for the constant $\Omega$ case in \S\ref{sec:Omega}
and constant $J$ in \S\ref{sec:J}. Conclusions are given in \S\ref{sec:Conclusions}.

\section{Static black holes \label{sec:Static}}

A non-rotating, neutral, asymptotically anti-de Sitter black hole has
line element
\[ d^2s = -f(r) dt^2 + f^{-1}(r)dr^2 + r^2 d\Omega^2, \]
with 
\beq \label{fdef}
f(r) = 1 -\frac{2 m}{r} - \frac{\Lambda}{3} r^2, 
\eeq
and $d\Omega^2=d\theta^2+\sin^2\theta d\phi^2$.
The horizon radius, $r_h$, is determined by the largest real
root of $f(r)=0$ giving 
\beq m= \frac {r_h}{2} \left(1  - \frac{\Lambda}{3} r_h^2 \right),
\label{Mass}
\eeq
which is the asymptotically AdS equivalent of the ADM mass, 
$M=m$, in the non-rotating case. 
Following \cite{KRT}, $M$ will be identified with the enthalpy,

\beq \label{Enthalpy}
H(S,P)=\frac {1} {2} \left(\frac {S}{\pi}\right)^{\frac 1 2} 
\left(1+\frac {8 P S} {3} \right),\eeq
where the Bekenstein-Hawking entropy is $1/4$ of the event horizon area 
\[ S=\pi r_h^2\]
and $P=-\frac{\Lambda}{8\pi}$ is the pressure ($G_N$ and $\hbar$ are set to one).  The temperature follows
either from the surface gravity 
\beq T=\frac{f'(r_h)}{4\pi}=\frac {\left(1-\Lambda r_h^2\right)} {4\pi r_h}
\label{eq:T_rLambda}\eeq 
using (\ref{fdef}) or 
from the thermodynamic relation
\beq T=\left.\frac{\partial H}{\partial S}\right|_P
=\frac{(1+8PS)}{4\sqrt{\pi S}}
\label{eq:T_SP}\eeq
using (\ref{Enthalpy}), these are the same formula written in geometric
and thermodynamic variables respectively.

The temperature has a minimum at $S=\frac{1}{8 P}$ when 
\beq T_{min}= \sqrt{\frac{2P}{\pi}}. \label{eq:T_min}\eeq
For $S<\frac{1}{8 P}$ the heat capacity 
\beq
C_P= T \left. \frac{\partial S}{\partial T}\right|_P = 
2S\left(\frac{8 P S +1}{8 P S -1}\right)
\eeq
is negative while for $S>\frac{1}{8 P}$ it is positive and it diverges at $T_{min}$.

The thermodynamic volume is the Legendre transform of the pressure, \cite{KRT} \cite{BPD1}, namely
\beq
V=\frac{\partial M}{\partial P}= \frac{4}{3\sqrt{\pi}}S(T,P)^{\frac 3 2}
=\frac{4\pi}{3} r_h^3.\label{eq:SchwarzschildVolume}
\eeq
This is a rather surprising result as there is no a priori reason 
for the thermodynamic volume to be related to a geometric volume.\footnote{It 
is not immediately clear how the geometric volume of a black hole might be defined, as
$r$ is a time-like co-ordinate and $t$ a space-like co-ordinate
for $r<r_h$. Inside the event horizon surfaces of constant $t$ have a 
time-dependent metric. Note that the entropy and the volume are not independent for a Schwarzschild black hole: this is not a pathology, they are independent for rotating black black holes
and the above formula can be obtained by taking the non-rotating limit
of the rotating case \cite{BPD2}.} Indeed when rotation is introduced
there is no such obvious relation \cite{CGKP}.

The Gibbs free energy is the Legendre transform of the enthalpy,
\[ G(T,P)= H-TS = \frac{1}{4} \sqrt{\frac{S(T,P)}{\pi}}\left( 1-\frac{8PS(T,P)}{3}\right)=\frac{r_h}{4}\left( 1+\frac{\Lambda}{3} r_h^2\right),\]
with
\beq
S(T,P)=\frac{\pi T^2 - P \pm T\sqrt{\pi^2 T^2 - 2 \pi P}}{8 P^2}
\eeq
(the heat capacity is positive for the plus sign, negative for the minus sign
and diverges when $P\rightarrow \pi T^2/2$).

We have the thermodynamic relation 
 \beq dG=-S dT + V dP \eeq
and, for AdS space-time, $G=0$ and so $S_{AdS}=V_{AdS}=0$ (this is the thermodynamic volume of AdS space-time, not a geometric volume). 
For a black hole, on the other hand,
the thermodynamic volume is given by (\ref{eq:SchwarzschildVolume}).

If $PS>\frac{3}{8}$ the Gibbs free energy of the black hole is negative
and thus lower than that of anti-de Sitter space-time, the former is 
then the more
stable thermodynamic configuration, while for $PS<\frac{3}{8}$ 
pure anti-de Sitter
space-time is the more stable and any black hole with $S<\frac{3}{8P}$
will tend to evaporate.  This is the Hawking-Page phase transition,
\cite{HawkingPage}, which occurs on the line $\Lambda r_h^2=-3$, or
\beq S=\frac{3}{8P} \qquad \Rightarrow \qquad 
P=\frac{3\pi}{8}T^2,\label{eq:TP}\eeq
when the two states can exist together as shown in the $P-T$ plane in 
figure \ref{fig:static_free_energy} below.  
(Similar phase diagram plots appeared in \cite{Altamirano} and are shown
here with a view to extending the analysis to constant angular velocity in the
next section.)

\begin{figure}[ht]
\centerline{\raise 1cm \hbox{\includegraphics[width=7cm,height=5cm]{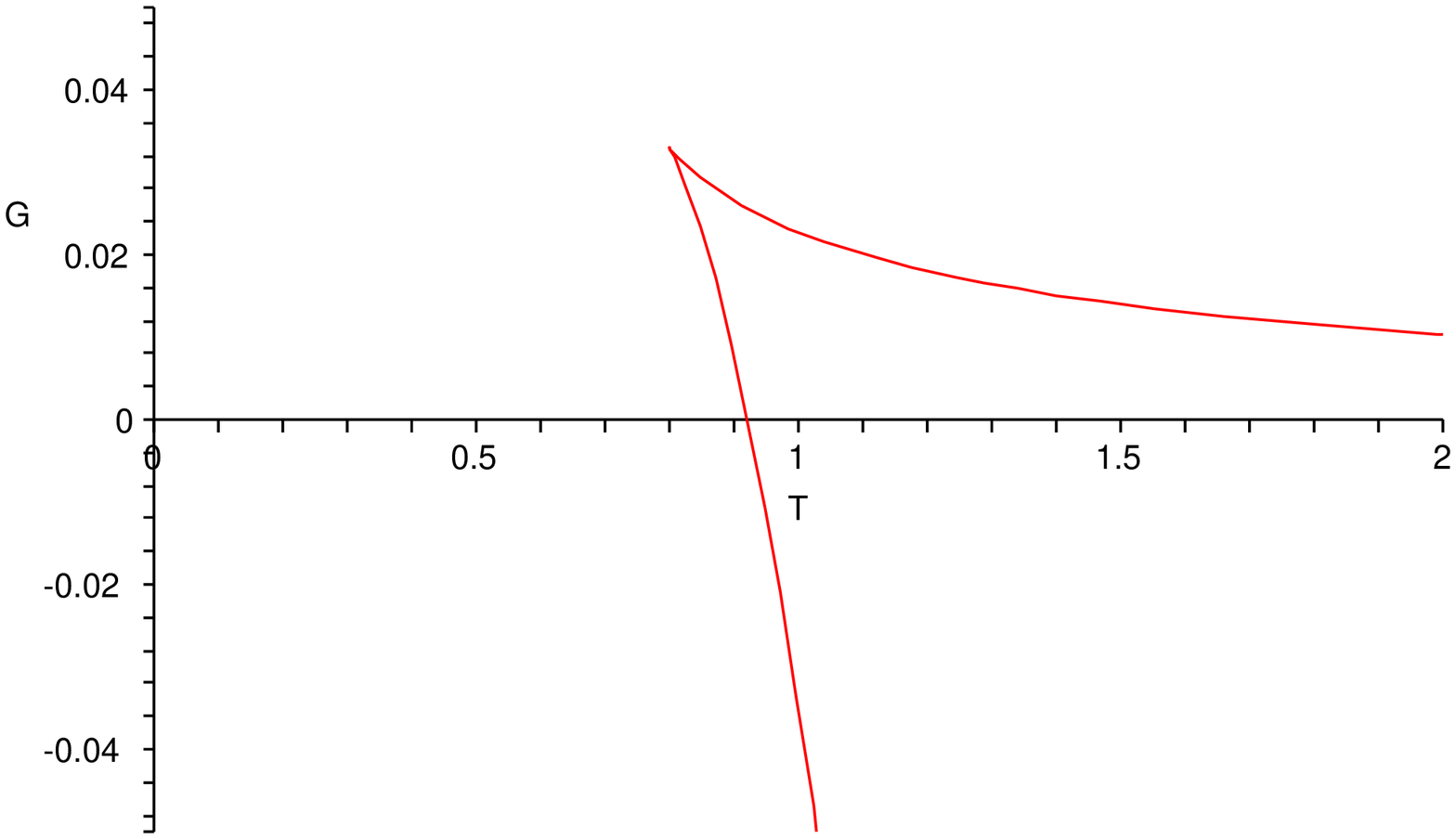}} 
 \hbox{\includegraphics[width=7cm]{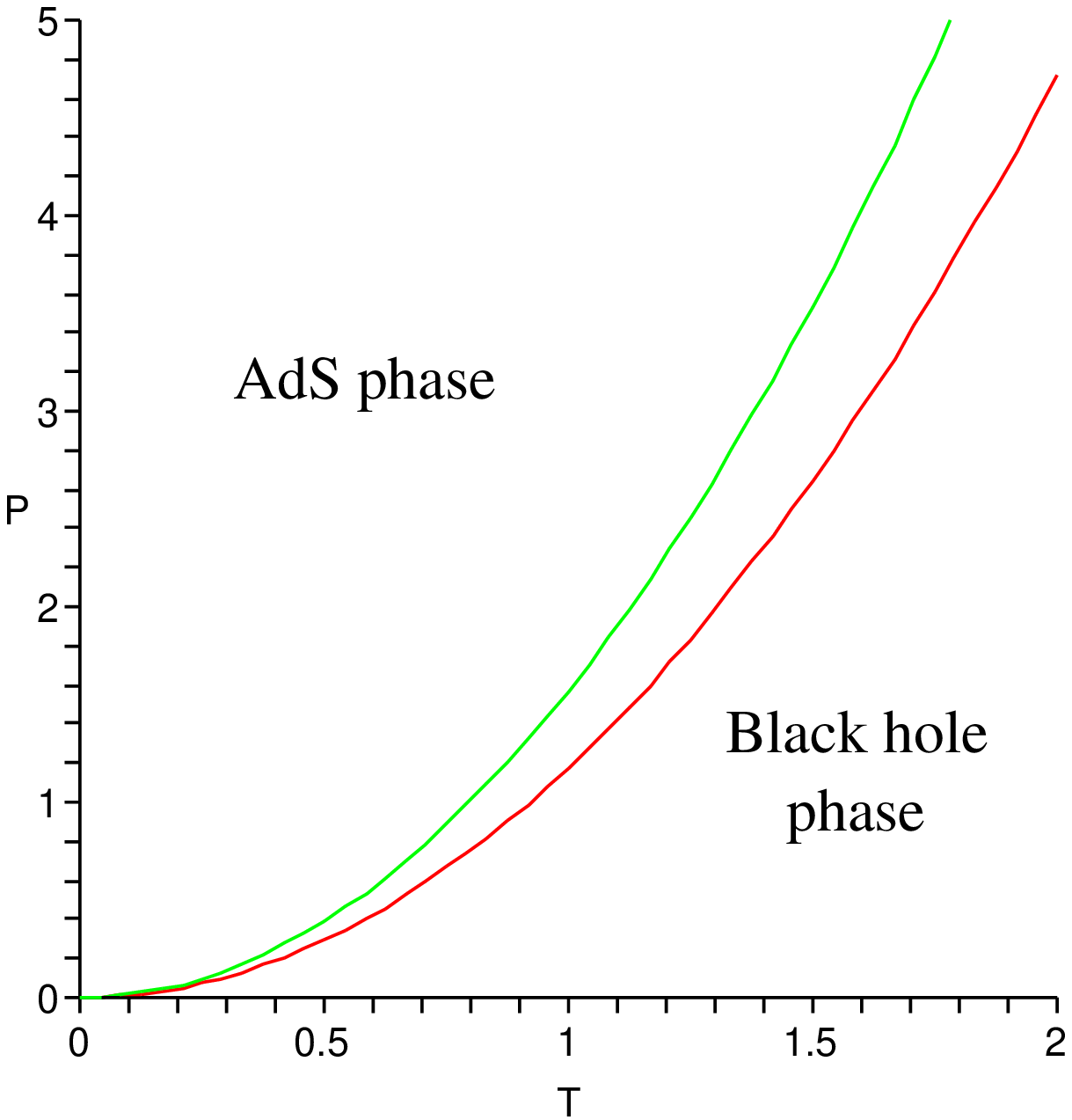}}}
\caption{\small Left: the black hole free energy, for static black holes, at fixed $P=1$, the heat capacity is negative on the upper branch, positive on the lower branch and diverges at the cusp.
The Hawking-Page temperature is where $G=0$.
Right: the co-existence curve of the Hawking-Page phase transition is the red (solid) line, the heat capacity diverges on the green (dotted) line.
The lower branch of the free energy tends to minus infinity on the $P=0$ axis.}  
\label{fig:static_free_energy}
\end{figure}

Clearly there is a jump in entropy,
\beq \Delta S = \frac{3}{8P}, 
\label{eq:DeltaS}
\eeq
when a black hole nucleates from pure AdS space time, energy must be supplied
to form the black hole at constant temperature and the latent heat is 
\beq L=T\Delta S. \label{eq:TDeltaS}\eeq
From the form of the co-existence curve in (\ref{eq:TP}) the latent heat is
\beq L=\frac{1}{\pi T}=\sqrt{\frac{3}{8\pi P}},\eeq
which is equal to the mass on the co-existence curve,
in the black hole phase.  The latent heat is non-zero for
any finite $T$ and goes to zero as $T\rightarrow\infty$, as though the system
were aiming for a second order phase transition but cannot reach it.
For asymptotically flat space-times, $P=0$ and the latent heat is infinite,
hence black holes will not spontaneously nucleate in Minkowski space.

The Clapeyron equation for static black holes follows from equating the Gibbs free energy of the
two phases at a point on the co-existence curve. 
For AdS space-time $G_{AdS}=0$, so in particular
\[ dG_{AdS}=0 \]
on the co-existence curve, while for the black hole
\[ d G_{BH} = -S_{BH}dT + V_{BH}dP.\]
The co-existence curve is defined by
$G_{AdS}=G_{BH}$, so
\bea 0 &=&d G_{AdS}- dG_{BH} = S_{BH} dT - V_{BH} dP \nonumber \\
\Rightarrow \frac{dP}{dT}&=&\frac{S_{BH}}{V_{BH}}=\frac{3}{4 r_h}. \eea
Since there is no black hole in pure AdS space-time we define
\[ \Delta V = V_{BH},\qquad \Delta S = S_{BH}\]
giving
\beq \frac{dP}{dT}=\frac{\Delta S}{\Delta V},\label{eq:Clapeyron}\eeq
which is the Clapeyron equation \cite{Callen}.
It is easily checked, using (\ref{eq:T_rLambda}) and (\ref{eq:TP}) directly, 
that indeed $\frac{dP}{dT}=\frac{3}{4 r_h}$.
The Clapeyron equation for static charged black holes, and its relation to Ehrenfest's
equations, was considered in \cite{BKR}.

\section{Rotating black holes \label{sec:Omega}}

In this section the analysis of the Hawking-Page phase transition 
is extended to rotating black holes in asymptotically AdS space times.
The ADM mass for a black hole rotating with angular
momentum $J$ can be expressed as a function of the entropy, the angular
momentum and the pressure \cite{CCK}
\beq \label{eq:CCKmass}
H(S,P,J):=
\frac {1}{2}\sqrt{\frac{\left( 1+\frac{8 P S}{3}  \right)
\left(S^2\left(1   + \frac{8 P S}{3}\right) + 
4 \pi^2 J^2\right)}
{\pi S}},
\eeq
which is again interpreted here as the enthalpy.

In asymptotically AdS space-times there can be rotating black holes 
that are in equilibrium with thermal radiation rotating at infinity, \cite{HHT-R} \cite{HawkingPage}.  The black hole is stable against decay
if its Euclidean action is less than that of pure AdS, which is zero,
so it is stable if the Euclidean action is negative.
The Euclidean action of the black hole, $I_E$, is a function of $T$, $P$ and the angular velocity,
$\Omega$, and is related to the Legendre transform of the ADM mass \cite{GPPI}, $I_E=\Xi/T$ with    
\beq \label{eq:Xi} \Xi(T,\Omega,P)=M-ST-J\Omega.\eeq
The Legendre transforms can be performed to obtain $\Xi$ explicitly.  First express the
temperature and the angular velocity as functions of entropy and angular 
momentum
\beq \label{CCKTemperature}
T= \left.\frac {\partial H}{\partial S}\right|_{J,P}=\frac {1}{8\pi H}\left[
\left(1  +\frac {8 P S}{3} \right)
\left(1 + 8 P S \right)
-4\pi^2 \left(\frac {J}{S}\right)^2\right]
\eeq
\beq \label{eq:CCKOmega}
\Omega = \left.\frac{\partial M}{\partial J}\right|_{S,P}
={2\,  {\pi }^{3/2}J} \sqrt{\frac{\left( 3+8\,PS \right)}
{S \left( 3\,{S}^{2}+8\,P{S}^{3}+12\,{\pi }^{2}{J}^{2} \right) }}\ .\eeq
While it is easy to invert the latter to write $J$ as a function of $\Omega$,
expressing $S$ explicitly as a function of $T$ requires the solution  of
an eighth order polynomial equation.
Nevertheless we can eliminate $J$ in favour of  $\Omega$,
\beq T(S,\Omega,P)= 
\frac{\left( 64\,{P}^{2}{S}^{2}\pi -24\,P{S}^{2}
{\Omega}^{2}+32\,\pi \,PS-6\,{\Omega}^{2}S+3\,\pi  \right)}
{4\pi\sqrt{\,S \left( 3+8\,PS \right)
\left( 3\,\pi +8\,\pi \,PS-3\,{\Omega}^{2}S \right)}
}.\label{eq:TempS}
\eeq
Solving for $S$ now only involves a quartic, but simpler is to use (\ref{eq:TempS}) 
in (\ref{eq:Xi}) to give
\beq \Xi(T(S,\Omega,P),\Omega,P)
={\frac {\sqrt{S}   \left(9\,\pi  +24\,P{S}^{2}{\Omega}^{2}-64\,{P}^{2}{S}^{2}\pi 
 \right) }{12 \pi \,\sqrt { \left( 3+8\,PS \right)\left( 3\,
\pi +8\,\pi \,PS-3\,{\Omega}^{2}S \right) }}},
\label{eq:Xi2}
\eeq
which, together with (\ref{eq:TempS}), gives $\Xi(T,\Omega,P)$
parametrically in terms of $S$.

The Hawking-Page phase transition is determined by the locus of points
where $\Xi=0$, {\it i.e.}
\beq \label{eq:HPCo-Existence}
S^2=\frac{9\pi}{8P(8\pi P - 3\Omega^2)}.\eeq
Note that 
\beq \Omega^2\le\frac{8\pi P}{3}\label{eq:Pless3}\eeq 
is a condition that must be imposed 
on $\Omega$ in order to ensure that the Einstein universe at infinity 
is not rotating faster than the speed of light \cite{HHT-R}.

\begin{figure}[ht] 
\centerline{\includegraphics[width=7cm]{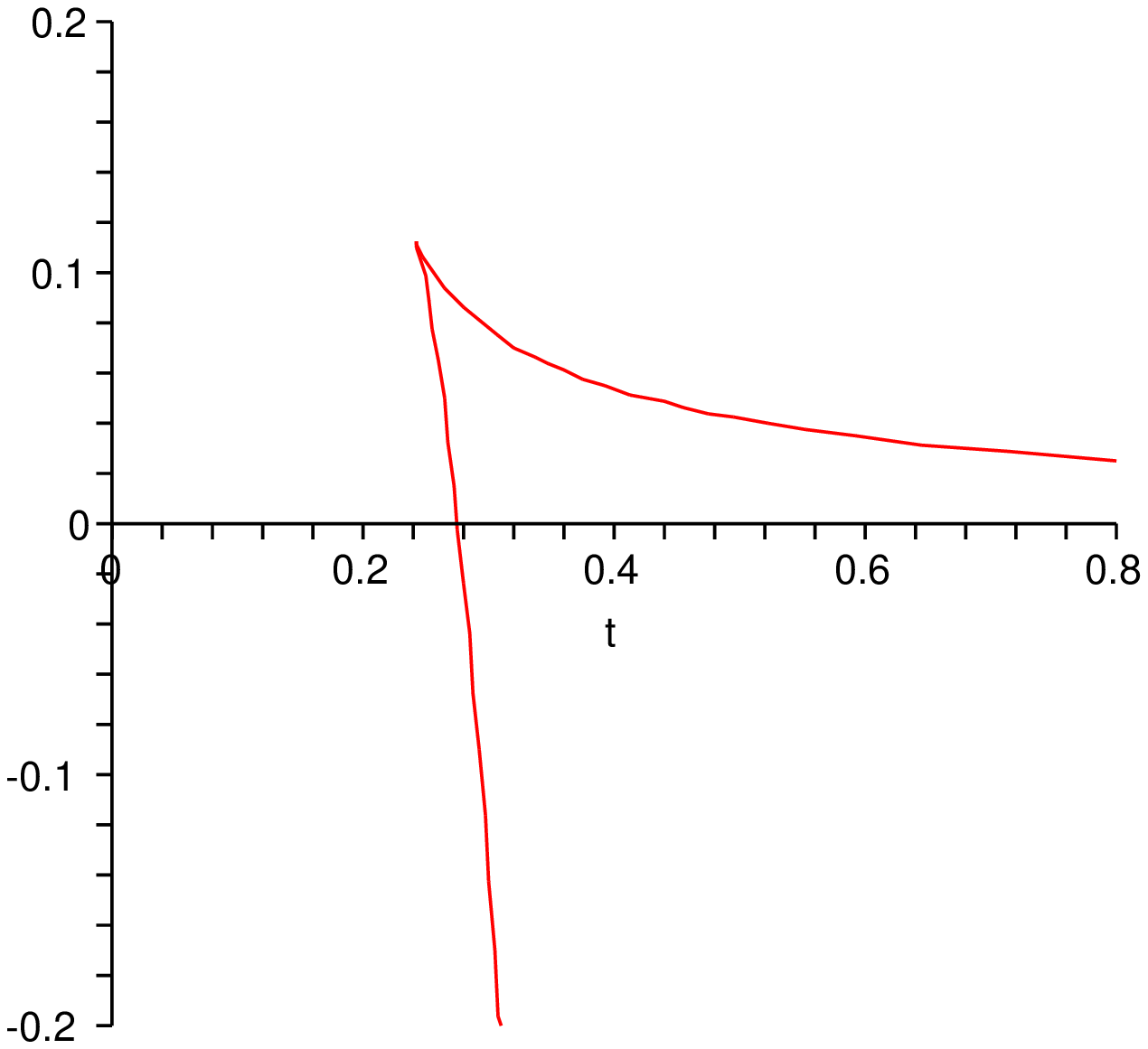}\includegraphics[width=7cm]{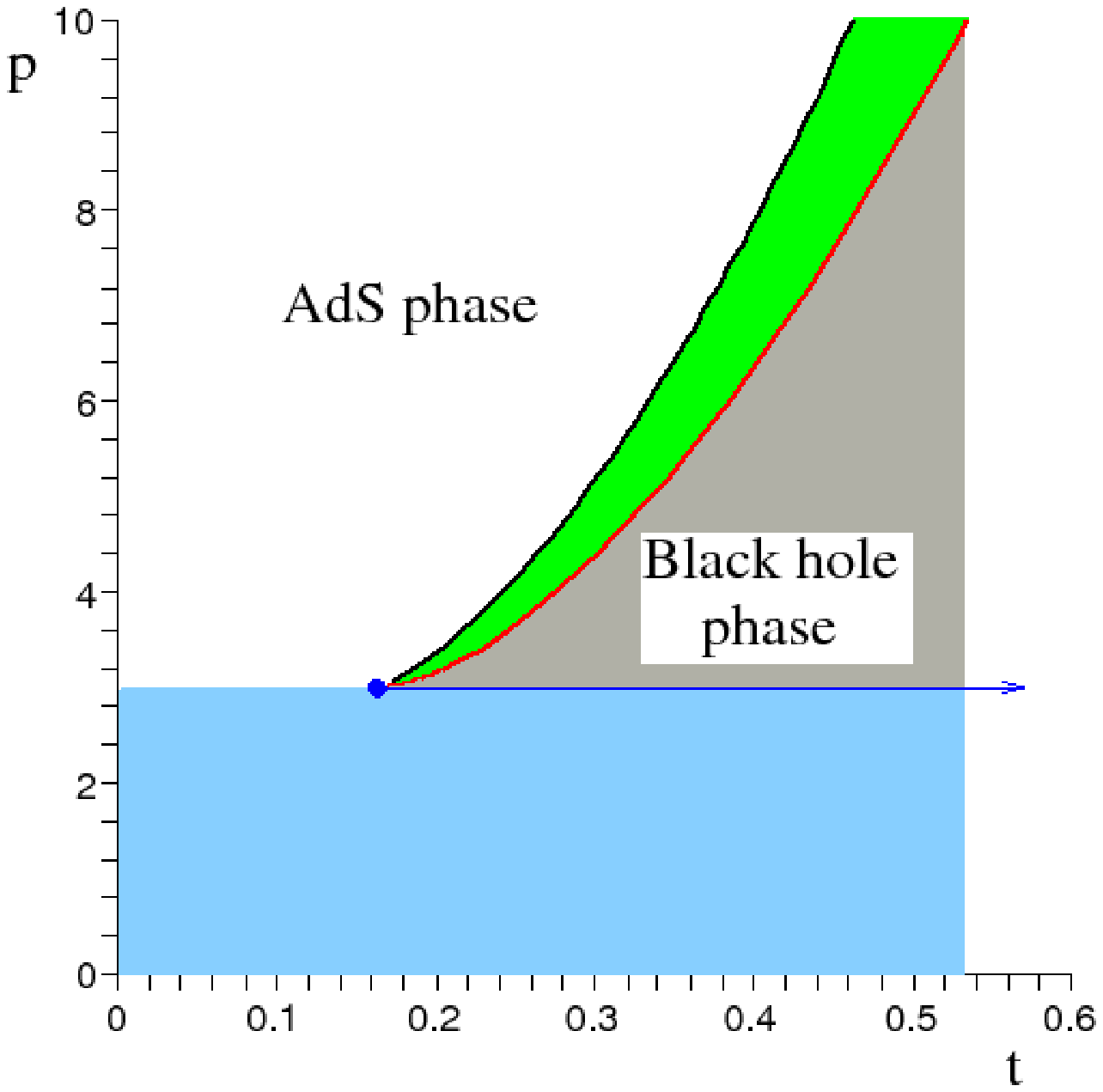}}
\caption{\small Left: the dimensionless
free energy for a rotating black hole as a function of the dimensionless variable $t$. 
The upper branch represents small black holes (negative heat capacity), the lower branch large black holes (positive heat capacity). Only negative $\Xi$ black holes are stable against Hawking-Page
decay. The vertical axis is the dimensionless combination $\Omega\, \Xi$
and the figure is drawn for $p=4$ (other values of $p>3$ move the position of the
cusp but the shape of the figure is the same).
Right: phase diagram in the $p-t$ plane. The constraint that the Einstein universe at infinity is not rotating faster than the speed of light imposes the condition $p\ge 3$. In the region below this line (blue) $\Xi_{BH}$ is single valued and positive. In the white region (upper left labelled \lq\lq AdS phase'') there are no black hole solutions (a black hole in this region would have negative entropy).  The right-hand boundary of the white region is the locus of
points on which the heat capacity diverges (black line), to the right of this there is
a wedge shaped region (green) in which $\Xi_{BH}$ has two branches and is positive on both branches. AdS is still the preferred phase in this region as
$0=\Xi_{AdH}<\Xi_{BH}$. In the grey region, labelled \lq\lq Black hole phase'' one branch of $\Xi_{BH}$ becomes negative and black holes are more stable than AdS.
The Hawking-Page phase transition occurs on the boundary between the green and the grey region (red line), where the lower branch of $\Xi_{BH}$ equals zero. 
}
\label{fig:Omega_free_energy}
\end{figure}

The free energy at a fixed pressure $P>\frac{3\Omega^2}{8\pi}$ 
is plotted as a function of temperature
in figure \ref{fig:Omega_free_energy},
using the dimensionless variables
\beq p=\frac{8\pi P}{\Omega^2}\ge 3,\qquad t=\frac{T}{\Omega} \qquad
\hbox{and} \qquad s= \frac {\Omega^2 S}{\pi}. 
\label{eq:ptsdef}\eeq
For $p>3$ there are two branches and the Hawking-Page temperature is determined
by the point where the lower branch cuts the $t$-axis.
As $p\rightarrow 3$ from above the lower branch becomes steeper until it 
disappears at $p=3$, the upper branch remaining and terminating at $T=\frac{\Omega}{2 \pi}$, where $\Xi=\frac{1}{4\Omega}$.  For $p<3$ there are black hole solutions for all positive $T$ with the black hole free energy $\Xi_{BH}$ a positive function decreasing monotonically with $T$.

Substituting (\ref{eq:HPCo-Existence}) in (\ref{eq:TempS}) gives 
an analytic expression for the co-existence curve $T(P)$ at fixed $\Omega$:
it has the parametric form
\bea\label{eq:Omega_HP}
p&=&\frac {3(s+\sqrt{s^2+4})}{2s},\\
t&=&\frac{\sqrt{2+\sqrt{s^2+4}}}{2\pi\sqrt{s}}\nonumber
\eea 
and is plotted in the right hand figure of figure 2 (red line).
It terminates at $p=3$ and the temperature of the black hole cannot go below 
$T=\frac{\Omega}{2\pi}$ for $p\ge 3$.  
Black holes are stable against decay to AdS
in the grey region bounded by the red and the horizontal blue lines, labelled \lq\lq Black hole phase''. 
In this region
$\Xi_{BH}$ is double valued, with a positive and a negative
branch, the negative branch being the more stable of the two 
and also more stable than AdS which has $\Xi_{AdS}=0$.  
Below the horizontal blue line the negative branch of $\Xi_{BH}$ disappears
and only the positive branch remains.  Across the blue line the negative branch
jumps discontinuously from minus infinity (black hole) to zero (AdS).\footnote{For finite jumps such phase transitions have been termed zeroth-order and
and it has been suggested that they could occur in 
superconductors and superfluids \cite{Maslov}.
Indeed it was proposed in \cite{HHT-R} that 
the horizontal line $p=3$ may be associated with a superfluid transition.}

When there is rotation present angular momentum contributes to the 
change in enthalpy across the co-existence curve, which is the latent heat.
For the Hawking-Page transition the latent heat is just the mass of the
black hole:
\begin{itemize}
 \item On the first order line, $\Xi_{BH}=M-TS-\Omega J=0$, hence
\[ L=M =T S + \Omega J.\]
Since the entropy and angular momentum vanish on the $AdS$ side of the transition the jumps
in entropy and angular momentum are $\Delta S=S$ and $\Delta J=J$, 
giving latent heat 
\beq L=T \Delta S + \Omega \Delta J =
\frac{16\pi^3 T^3}{(4\pi^2 T^2 -\Omega^2)^2},
\label{eq:latentOmega}
\eeq
which diverges at the critical point\footnote{This is a critical point
in the original sense of the phrase, it is necessary to tune both $T$ and $P$ very carefully to access this point. A critical point in this sense is a separate
concept to the notion of zero latent and a ``second order'' phase transition.} $T=\frac{\Omega}{2\pi}$,
$P=\frac{3\Omega^2}{2\pi}$.

Of course $M=TS+\Omega J$ is not a general formula, it only holds on the
co-existence curve and gives and gives a 
quick and convenient way of determining the curve. It can be combined with Smarr relation,
\[ M=2(TS+\Omega J - PV),\]
to give $M=2P V$, or
\beq \rho=2 P
\label{eq:PVHP}\eeq
with $\rho=\frac M V$, on the co-existence curve.  

The Clapeyron equation can be checked using 
thermodynamic volume, $V=\left.\frac{\partial H}{\partial P}\right|_{S,J}$
first calculated in \cite{CGKP}.
Written as a function of $\Omega$ it reads
\beq
V={\frac {2S^{\frac 3 2 }\left( 16\,\pi \,PS-3\,{\Omega}^{
2}S+6\,\pi  \right)}{3\pi\sqrt {{ { 
\left( 3+8\,PS \right)}{(3\,\pi +8\,\pi \,PS-3\,{\Omega}^{2}S)}
}}}}. \label{eq:Volume}\eeq
On the co-existence curve (\ref{eq:HPCo-Existence}) this is
\[ V=\frac{2\pi}{3\Omega^3}\sqrt{s^3(2+\sqrt{s^2+4})}\;.\]
Again, as the thermodynamic volume of AdS without a black hole vanishes,
$\Delta V=V$ and 

\[\frac{\Delta S}{\Delta V} = \frac{3\Omega}{2\sqrt{s(2+\sqrt{s^2+4})}},
\]
and this is indeed equal to $\frac{dP}{dT}$ along the co-existence curve,
from (\ref{eq:Omega_HP}).

Another way of writing this is to use (\ref{eq:PVHP}) to give
\beq \label{eq:dTdP}
\frac{d P}{d T}= \frac{S}{V}=\frac{2PS}{M}
\eeq
hence, at a given value of $\Lambda$, the slope of the co-existence
curve is determined by the entropy per unit mass of the black hole.

\item 
The Clapeyron equation does not hold across the zeroth order transition
line in figure 2 (blue), because $\Xi_{BH}\ne 0$ there.
In fact $\Xi_{BH}<0$ on the blue line and the free energy jumps across
this line


\end{itemize}

The free energy $\Xi_{BH}$ is plotted in figure \ref{fig:Omega_free_energy} as a function of the dimensionless variable $t$ in (\ref{eq:ptsdef}), with $p=4$.  There is a cusp at the
minimum temperature $t=\frac{1}{2\pi}$, where the heat capacity and the latent
heat diverge, but no phase transition.

\section{The Caldarelli, Cognola and Klemm phase transition \label{sec:J}}

It was found in \cite{CCK} that there is a phase transition
for asymptotically AdS rotating black holes
with fixed angular momentum, $J$, between small and large black holes, the CCK phase transition.
The phase structure associated with constant $J$ transitions has been more
extensively studied from the point of view of varying $P$ than the constant
$\Omega$ case \cite{Altamirano}. It is more like the familiar liquid-gas transition than the constant $\Omega$ case and in higher dimensions there can be a triple point associated with three different black hole phases \cite{Altamirano}.
There is no Hawking-Page phase transition for constant $J$, because AdS with no black hole cannot have non-zero $J$. 
The results in this section are not new and
are included for completeness and comparison to the constant $\Omega$ case in section \S\ref{sec:Omega}.

The thermodynamic form of the mass for the asymptotically AdS Kerr metric is (\ref{eq:CCKmass}), the temperature is (\ref{CCKTemperature}) 
and the thermodynamic volume (\ref{eq:Volume}) is, in terms of $J$,
\beq\label{eq:ThermodynamicVolume}
V= \left.\frac {\partial M}{\partial P}\right|_{S,J}=
\frac{2}{3 \pi M}\left\{S^2\left(1 + \frac{8 P S}{3}\right)+ 
2\pi^2 J^2 \right\}.
\eeq
This is greater than (or equal to when $J=0$)
the na{\"\i}ve geometric result \hfill\break
$V=\frac{4\pi}{3}\left(\frac{S}{\pi}\right)^{\frac 3 2}$, 
as first observed in \cite{CGKP}.
 
\begin{figure}[ht]
\centerline{\includegraphics[width=7cm]{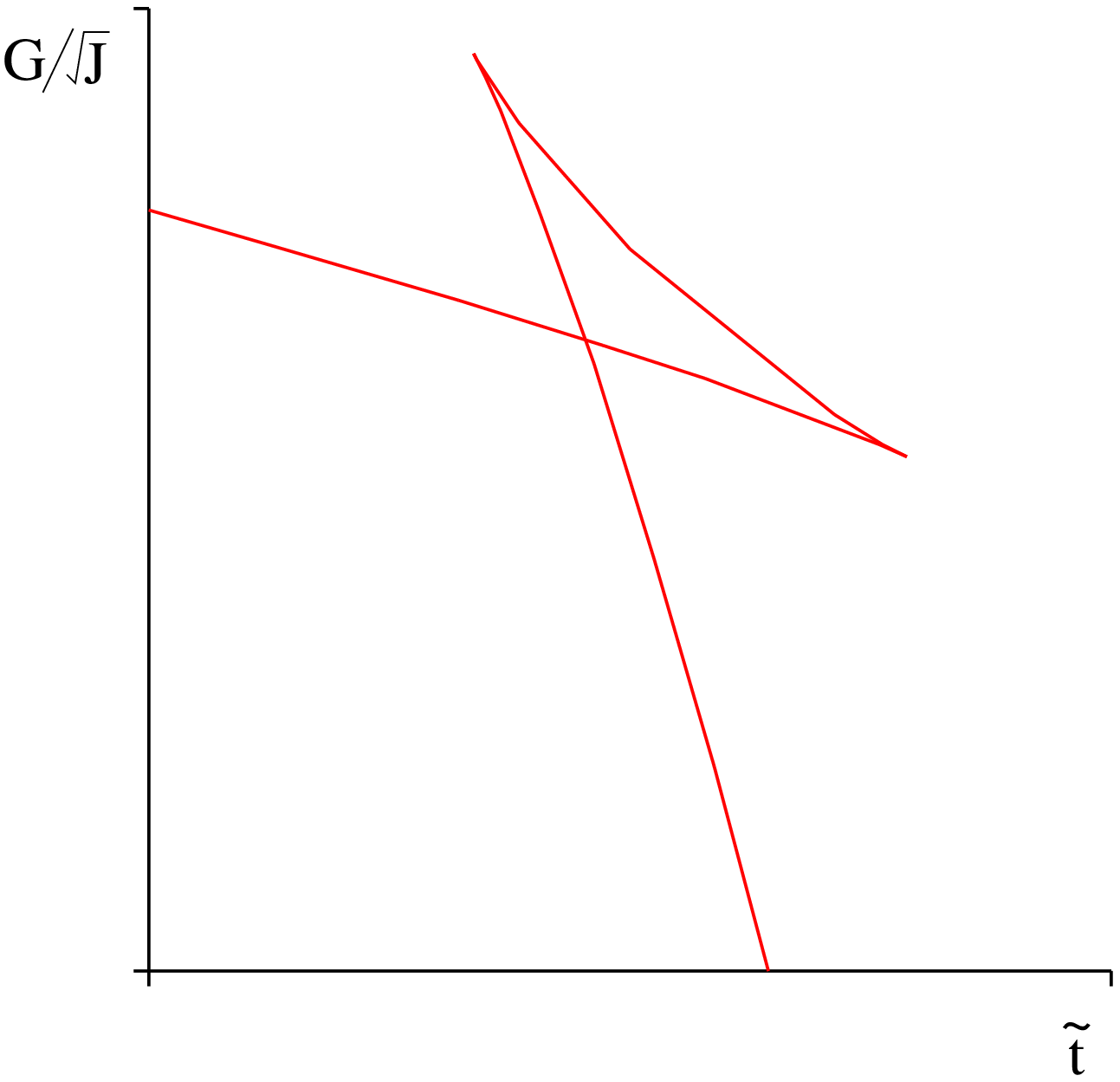}\includegraphics[width=7cm]{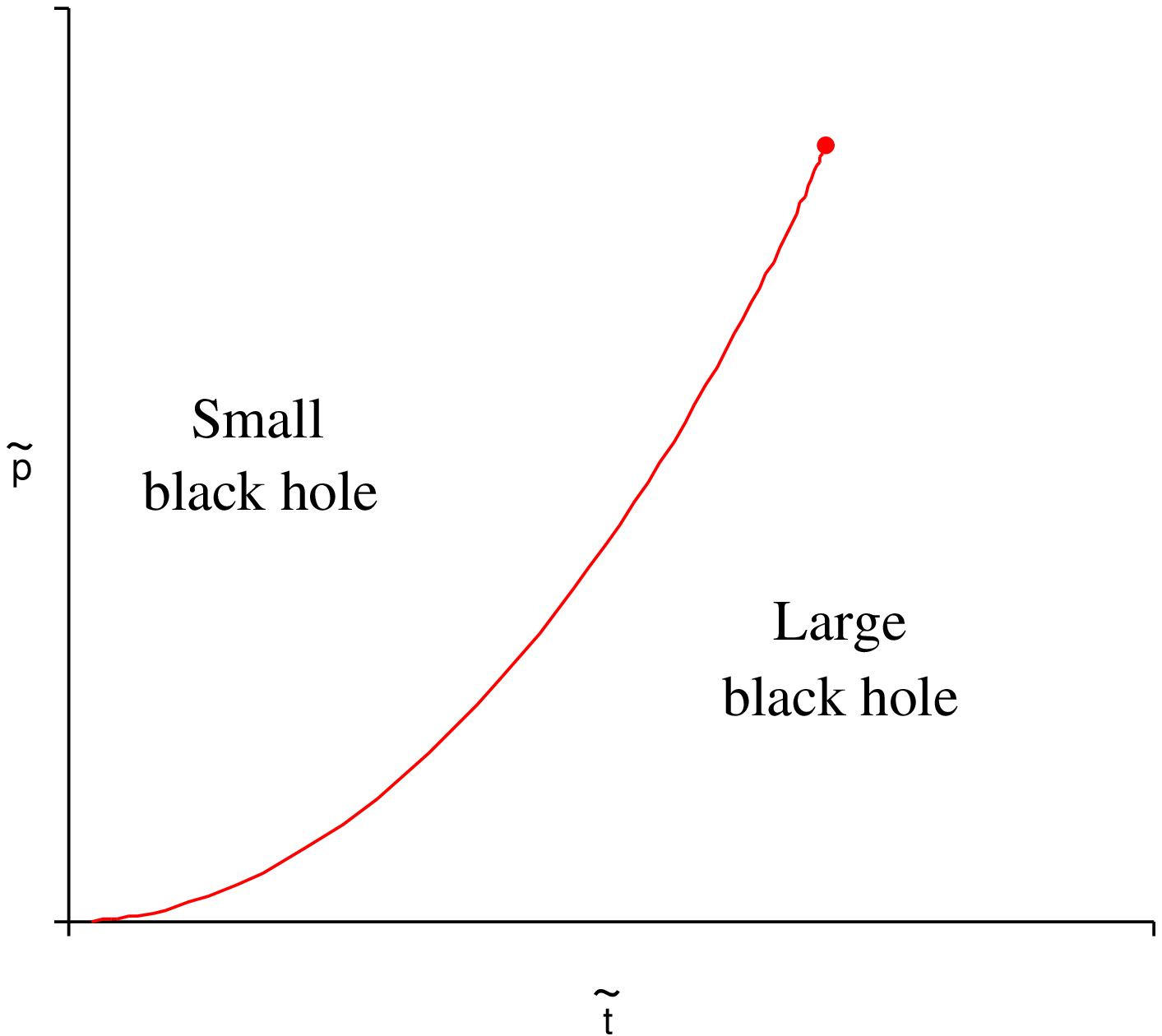}}
\caption{\small Left: Gibbs free energy, $G/\sqrt{J}$, as a function of $\t$ 
for a black hole at constant $J$ with $\p= <\p_c$.
Right:  co-existence curve at fixed $J$, in the $\p-\t$ plane. The curve ends at the
critical point $\t_c=0.0417$, $\p_c=0.144$.}
\label{fig:J_free_energy}
\end{figure}

In the $P-V$ plane the CKK transition mimics the van der Waals gas-liquid
phase transition very closely.  Below a critical temperature $T_c$ 
the transition is first order, culminating at a second order transition at
$T_c$, in the same universality class as the van der Waals transition (with
mean field exponents, \cite{GKM} \cite{PdV}) and no transition for $T>T_c$.
The free energy
\beq G(T,P)=M(S,P)-S T
\eeq
is plotted in figure \S\ref{fig:J_free_energy}, as a function of $T$ at fixed $P$, together with the co-existence curve in the $P-T$ plane. 

Just like the van der Waals equation of state there is a line
a line of second order phase transitions, but here between large and small black holes,
terminating at critical point where there is a second order phase transition.

The heat capacity at constant $J$ and $P$ is
\beq
C_{J,P}=T\left.\frac{\partial S}{\partial T}\right|_{J,P}
\eeq
The full expression is not very illuminating and we shall focus on the spinodal
curve at constant $J$, the locus of points in the $S-P$ plane where
the heat capacity diverges.  In terms of the dimensionless variables
\[\p=16\pi P J, \qquad \s= \frac{S}{2\pi J}  \]
the spinodal curve is given by a quartic polynomial in $\p$,
\bea
&&\kern -30pt
{\p}^{4}{\s}^{8}+4\,{\s}^{5} \left( 2\,{\s}^{2}+3 \right) {\p}^{3}+18\,{\s}^{4}\left( {\s}^{2}+5 \right) {\p}^{2}+36\,\s \left( 6\,{\s}^{2}+1
 \right) \p+162\,{\s}^{2}+81-27\,{\s}^{4}\nonumber \\
&=&0.\nonumber 
\eea

\begin{figure}[ht]
\centerline{\includegraphics[width=12cm]{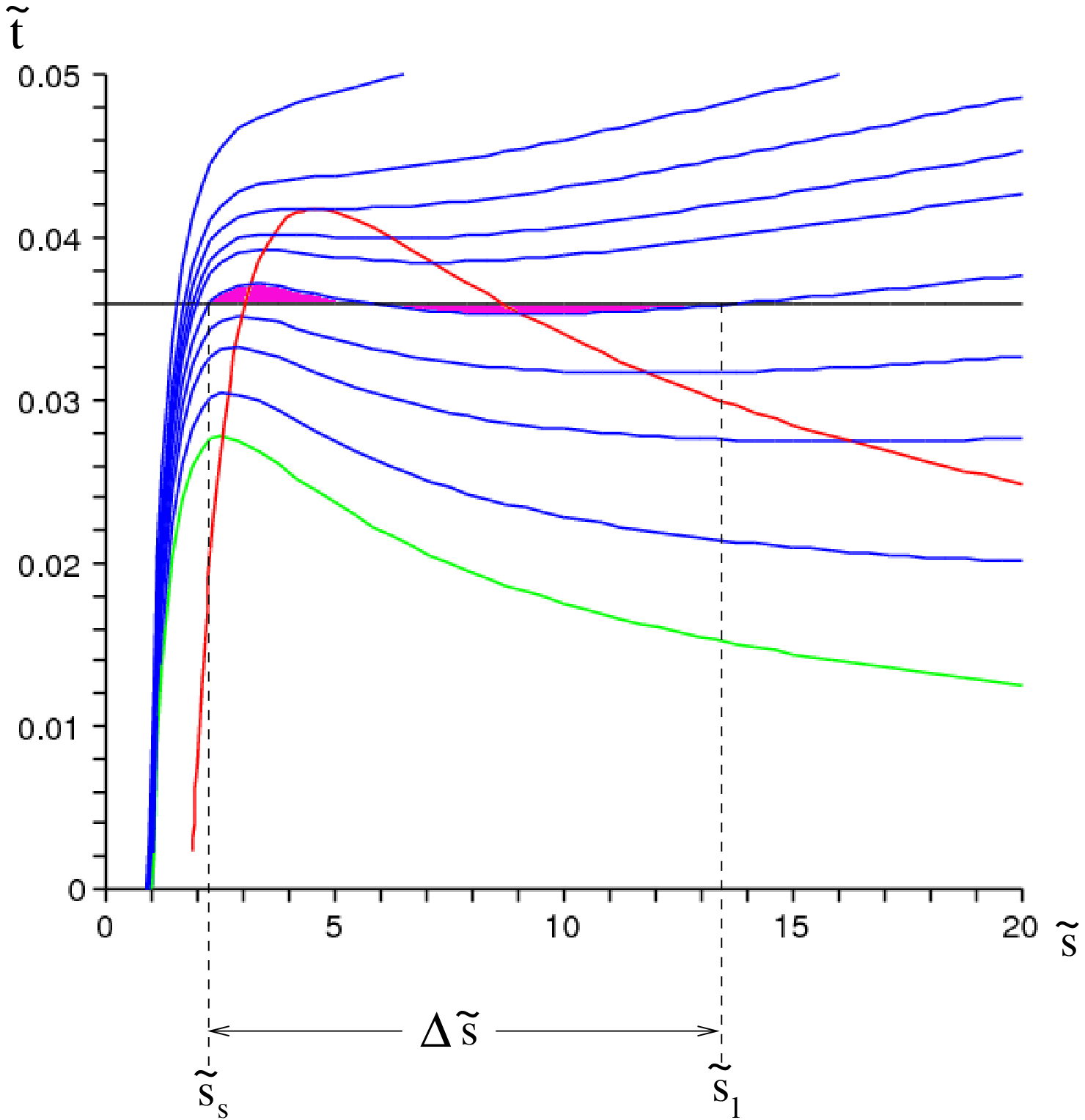}}
\caption{\small Curves of constant $\p$ in the $\s-\t$ plane.
The green curve is $\p=0$. The red curve is the spinodal curve. The temperature of the phase transition
is calculated using the Maxwell equal area rule.}
\label{fig:T-S_diagram}
\end{figure}

Defining a dimensionless temperature $\t=T\sqrt{J}$, curves of constant $\p$ are
plotted in the $\s-\t$ plane in figure \ref{fig:T-S_diagram}.

The spinodal curve is the red curve and the negative slope of the lines
of constant $P$ under the spinodal curve is an unstable region, where the
heat capacity is negative.  The temperature of the large-small black hole
phase transition can in principle be obtained from the Maxwell equal area rule:
the Gibbs free energy, at constant $J$,
\[ G(T,P,J)=M(S,P,J)-TS \]
should be the same in both phases, $G_l=G_s$. \

 At constant $P$
\[ dG=-S dT + V dP =-S dT,\]
so 
\bea 0 = G_l - G_s = \int_{S_s}^{S_l}dG &=& -\int_{S_s}^{S_l} S d T
=-[ST]^l_s+\int_{S_s}^{S_l} T(S)\, d S\nonumber \\
\Leftrightarrow \qquad \int_{S_s}^{S_l}T(S)\,d S-(S_l-S_s)T   &=&0.\nonumber \eea
The last expression determines 
the transition temperature, $T=T(S_l)=T(S_s)$, by demanding that the filled
area in figure \ref{fig:T-S_diagram} is zero.   
The two phases co-exist on a line in the $T-S$ plane, terminating at the 
critical point where the two sizes are equal
(the Maxwell equal area rule for the first order Hawking-Page transition, 
associated with static AdS-Schwarzschild, was investigated in the $P-V$ plane in \cite{SS}).

One must however be careful about applying ordinary thermodynamic intuition to black holes.  
In the $S-T$ plane the segment of an isotherm on which $T$ decreases with
$S$ corresponds to negative heat capacity and signals
an instability.  The Maxwell rule states that, for the liquid-gas phase transition,  such an isotherm should be replaced by
one which is horizontal between the two extremes of pure gas and pure liquid,
the horizontal section of the isotherm being a mixture of liquid and gas in linear proportion to the distance to its two end points 
\cite{Callen} (the argument is usually given in the $P-V$ plane, when
compressibility replaces heat capacity, but it is essentially the same).  
There does not appear to be any such interpretation for black holes: a classical black hole solution is not a combination of large and small
black holes, it is either one or the other and there does not seem to be any
simple way in which the horizontal
section of the constant $P$ curve in figure \ref{fig:T-S_diagram} can be thought of 
as a \lq\lq mixture'' of black holes with different entropies, as there is
only one black hole.
The negative heat capacity in the unstable region of figure \ref{fig:T-S_diagram},
and the negative compressibility \cite{Compressibility},
may be a feature that one must
live with, just like the negative heat capacity of asymptotically flat
Schwarzschild black holes.

The Clapeyron equation $\frac {d P}{d T} = \frac{\Delta S}{\Delta V}$, follows from standard reasoning, that the Gibbs free energy
of the two phases agrees across on the co-existence curve,
$G_l(T,P)=G_s(T,P)$,
but an explicit verification for the CCK transition would require 
solving a high order polynomial by numerical computation.

\section{Conclusions \label{sec:Conclusions}}

Phase transitions for asymptotically AdS Kerr black holes have been analysed
in thermodynamic variables, viewing the (negative) 
cosmological constant as a pressure and taking its thermodynamically conjugate
variable to be a volume.
For non-rotating black holes the analytic form of the Hawking-Page
transition line in the $P-T$ plane is quadratic (\ref{eq:TP}) and the latent
heat is inversely proportional to the temperature.
At constant $\Omega$ the transition line is given parametrically in equations
(\ref{eq:ptsdef}) and (\ref{eq:Omega_HP})
while the latent heat is (\ref{eq:latentOmega}) and is proportional to 
the mass per unit entropy.  The Clapeyron equation has been explicitly
checked to hold true, which is a consistency check on the thermodynamic
interpretation of $\Lambda$ presented here.
Analytic expressions are harder to obtain for constant $J$, but
similar concepts apply.

From the AdS/CFT perspective the constant $\Omega$ case is 
relevant for rotating superfluids in the boundary conformal field theory. 
The main difference between the phase diagrams when $\Omega=0$ and
$\Omega\ne 0$ is that the lower boundary of the black hole phase
is at $P=0$ in the former case, and cannot be crossed for positive $P$,
whereas it as at $P>0$ in the latter case, corresponding to Bose condensation
on the boundary conformal field theory \cite{HHT-R}.

The phase diagram in higher dimensions is more complicated, due to 
the presence of more than one angular momentum, but is all the richer for that:
the constant $J$ case exhibiting triple points \cite{Triple} and re-entrant phase transitions \cite{Reentrant}.  It would be interesting to explore the
constant $\Omega$ phase diagram in higher dimensions in more detail.  
Bose condensation of vortices at a critical 
angular velocity for example could be a new phase when $\Omega\ne 0$.
 
There is much still to discover in this new picture
of black hole thermodynamics.

This research was supported in part by Perimeter Institute for Theoretical
Physics.  Research at Perimeter Institute is supported by the 
Government of Canada through Industry Canada and by the Province of Ontario
through the Ministry of Economic Development and Innovation.

\begin{appendix}

\end{appendix}

\end{document}